\documentclass[aps,prl,preprint,tightenlines,superscriptaddress,showpacs,byrevtex]{revtex4}

\usepackage{graphicx} 

\renewcommand{\bar}[1]{\overline{#1}{}}

\newcommand{\de}{\Delta E}
\newcommand{\dm}{\Delta m}
\newcommand{\dt}{\Delta t}
\newcommand{\mbc}{M_{\rm bc}}
\newcommand{\bb}{B{\bar B}}
\newcommand{\qq}{q{\bar q}}
\newcommand{\ks}{K^0_S}

\newcommand{\db}{\bar{D}^0}
\newcommand{\ds}{\bar{D}^{*0}}
\newcommand{\bdnpn}{\bar{B}^0\to D^0 h^0}
\newcommand{\bdspn}{\bar{B}^0\to D^{*0} h^0}

\newcommand{\kspipi}{\ks\pi^+\pi^-}
\newcommand{\dnkspipi}{D^0\to\ks\pi^+\pi^-}

\newcommand{\bdbkspipi}{\bar{B}^0\to D[ \kspipi] h^0}
\newcommand{\fq}{\ensuremath{q}}
\newcommand{\sinphi}{\sin 2\phi_1  = 0.726 \pm 0.037}
\newcommand{\phires}{\phi_1=(16\pm21\pm 11)^\circ}
\newcommand{\phicl}{-30^\circ<\phi_1<62^\circ}

\begin{document}

\preprint{\vbox{ \hbox{   }
    \hbox{BELLE-CONF-0546}
    \hbox{LP2005-181}
    \hbox{EPS05-521}
}}

\title{\quad\\[0.5cm] \boldmath Measurement of $\phi_1$
  using $\bdbkspipi$ 
}
\affiliation{Aomori University, Aomori}
\affiliation{Budker Institute of Nuclear Physics, Novosibirsk}
\affiliation{Chiba University, Chiba}
\affiliation{Chonnam National University, Kwangju}
\affiliation{University of Cincinnati, Cincinnati, Ohio 45221}
\affiliation{University of Frankfurt, Frankfurt}
\affiliation{Gyeongsang National University, Chinju}
\affiliation{University of Hawaii, Honolulu, Hawaii 96822}
\affiliation{High Energy Accelerator Research Organization (KEK), Tsukuba}
\affiliation{Hiroshima Institute of Technology, Hiroshima}
\affiliation{Institute of High Energy Physics, Chinese Academy of Sciences, Beijing}
\affiliation{Institute of High Energy Physics, Vienna}
\affiliation{Institute for Theoretical and Experimental Physics, Moscow}
\affiliation{J. Stefan Institute, Ljubljana}
\affiliation{Kanagawa University, Yokohama}
\affiliation{Korea University, Seoul}
\affiliation{Kyoto University, Kyoto}
\affiliation{Kyungpook National University, Taegu}
\affiliation{Swiss Federal Institute of Technology of Lausanne, EPFL, Lausanne}
\affiliation{University of Ljubljana, Ljubljana}
\affiliation{University of Maribor, Maribor}
\affiliation{University of Melbourne, Victoria}
\affiliation{Nagoya University, Nagoya}
\affiliation{Nara Women's University, Nara}
\affiliation{National Central University, Chung-li}
\affiliation{National Kaohsiung Normal University, Kaohsiung}
\affiliation{National United University, Miao Li}
\affiliation{Department of Physics, National Taiwan University, Taipei}
\affiliation{H. Niewodniczanski Institute of Nuclear Physics, Krakow}
\affiliation{Nippon Dental University, Niigata}
\affiliation{Niigata University, Niigata}
\affiliation{Nova Gorica Polytechnic, Nova Gorica}
\affiliation{Osaka City University, Osaka}
\affiliation{Osaka University, Osaka}
\affiliation{Panjab University, Chandigarh}
\affiliation{Peking University, Beijing}
\affiliation{Princeton University, Princeton, New Jersey 08544}
\affiliation{RIKEN BNL Research Center, Upton, New York 11973}
\affiliation{Saga University, Saga}
\affiliation{University of Science and Technology of China, Hefei}
\affiliation{Seoul National University, Seoul}
\affiliation{Shinshu University, Nagano}
\affiliation{Sungkyunkwan University, Suwon}
\affiliation{University of Sydney, Sydney NSW}
\affiliation{Tata Institute of Fundamental Research, Bombay}
\affiliation{Toho University, Funabashi}
\affiliation{Tohoku Gakuin University, Tagajo}
\affiliation{Tohoku University, Sendai}
\affiliation{Department of Physics, University of Tokyo, Tokyo}
\affiliation{Tokyo Institute of Technology, Tokyo}
\affiliation{Tokyo Metropolitan University, Tokyo}
\affiliation{Tokyo University of Agriculture and Technology, Tokyo}
\affiliation{Toyama National College of Maritime Technology, Toyama}
\affiliation{University of Tsukuba, Tsukuba}
\affiliation{Utkal University, Bhubaneswer}
\affiliation{Virginia Polytechnic Institute and State University, Blacksburg, Virginia 24061}
\affiliation{Yonsei University, Seoul}
  \author{K.~Abe}\affiliation{High Energy Accelerator Research Organization (KEK), Tsukuba} 
  \author{K.~Abe}\affiliation{Tohoku Gakuin University, Tagajo} 
  \author{I.~Adachi}\affiliation{High Energy Accelerator Research Organization (KEK), Tsukuba} 
  \author{H.~Aihara}\affiliation{Department of Physics, University of Tokyo, Tokyo} 
  \author{K.~Aoki}\affiliation{Nagoya University, Nagoya} 
  \author{K.~Arinstein}\affiliation{Budker Institute of Nuclear Physics, Novosibirsk} 
  \author{Y.~Asano}\affiliation{University of Tsukuba, Tsukuba} 
  \author{T.~Aso}\affiliation{Toyama National College of Maritime Technology, Toyama} 
  \author{V.~Aulchenko}\affiliation{Budker Institute of Nuclear Physics, Novosibirsk} 
  \author{T.~Aushev}\affiliation{Institute for Theoretical and Experimental Physics, Moscow} 
% ...
\collaboration{The Belle Collaboration}

\begin{abstract} 
  \noindent
We present a preliminary measurement of the angle $\phi_1$ of 
the CKM Unitarity Triangle using time-dependent analysis of
$D\to\kspipi$ decays produced in neutral $B$ meson decay to a neutral
$D$ meson and a light meson ($B\to D h$).
The method allows one to directly extract the value of $2\phi_1$.
The ambiguity between $2\phi_1$ and $\pi-2\phi_1$
in the measurement of $\sin 2\phi_1$ can then be resolved. We obtain 
$\phires$. The 95\% CL region including systematic error is
$\phicl$, thus ruling out the second solution from
$\sinphi$ at the 97\% CL.
\end{abstract}


\maketitle

{\renewcommand{\thefootnote}{\fnsymbol{footnote}}}
\setcounter{footnote}{0}


Precise determinations of the Cabibbo-Kobayashi-Maskawa (CKM) 
matrix elements~\cite{ckm} are important to 
check the consistency of the Standard Model and search for new physics.
The value of $\sin 2\phi_1$, where $\phi_1$ is one of the angles of 
the Unitarity Triangle~\cite{pdg_review}, 
is now measured with high precision: 
$\sinphi$~\cite{sin2phi1}. 

A new technique based on the analysis of $\bdbkspipi$ has been 
suggested recently~\cite{bgk}.
Here we use $h^0$ to denote a light neutral meson, such as 
$\pi^0$, $\eta$ or $\omega$.
The neutral $D$ meson is reconstructed in the $\kspipi$ mode,
whose amplitude content is well known.


Consider a neutral $B$ meson,
which is known to be a $\bar{B}^0$ at time $t_{\rm tag}$
(for experiments operating at the $\Upsilon(4S)$ resonance,
such knowledge is provided by tagging the flavour of the other $B$ meson
in the $\Upsilon(4S) \to B\bar{B}$ event).
At another time, $t_{\rm sig}$,
the amplitude content of the $B$ meson is given by
\begin{equation}
  \label{eq:b0bar_evo}
  \left| \bar{B}^0(\dt) \right> = 
  e^{-\left|\dt\right|/2\tau_{B^0}}
  \left(
    \left| \bar{B}^0 \right> \cos(\dm\dt/2) - 
    i \frac{p}{q} \left| B^0 \right> \sin(\dm\dt/2)
  \right),
\end{equation}
where $\dt = t_{\rm sig} - t_{\rm tag}$, 
$\tau_{B^0}$ is the average lifetime of the $B^0$ meson,
$\dm$, $p$ and $q$ are parameters of $B^0$-$\bar{B}^0$ mixing
($\dm$ gives the frequency of $B^0$-$\bar{B}^0$ oscillations,
while the eigenstates of the effective Hamiltonian in the 
$B^0$-$\bar{B}^0$ system are $\left| B_\pm \right> = 
p \left| B^0 \right> \pm q \left| \bar{B}^0 \right>$).
Here we have assumed $CPT$ invariance and 
neglected terms related to the lifetime difference of neutral $B$ mesons.

\begin{figure}
  \includegraphics[width=0.5\textwidth]{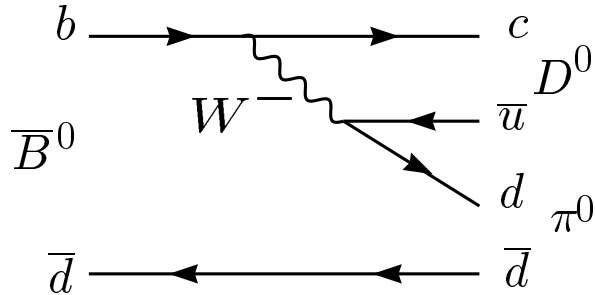}
  \caption{
    \label{diag_fav}
    Diagram for the dominant colour-suppressed amplitude 
    for $\bar{B}^0 \to D\pi^0$.
  }
\end{figure}

The $B\to D h$ decay is dominated by the CKM favored $b\to c\bar{u}d$
diagram as shown in Fig.~\ref{diag_fav} with roughly a 2\% contribution
from the CKM suppressed $b\to u\bar{c}d$ diagram. Ignoring the latter, 
a neutral $D$ meson produced in a $\bar{B}^0$ decay is a $D^0$, 
while that produced in a $B^0$ decay is a $\bar{D}^0$.
The $D$ meson state produced at time 
$\dt$ is then given by the following admixture:
\begin{equation}
  \label{eq:d_from_b0bar_evo}
  \left| \tilde{D}_{\bar{B}^0}(\dt) \right> = 
  \left| D^0 \right> \cos(\dm\dt/2) - 
  i \frac{p}{q} \xi_{h^0} (-1)^l \left| \bar{D}^0 \right> \sin(\dm\dt/2),
\end{equation}
where we use $\xi_{h^0}$ to denote the $CP$ eigenvalue of $h^0$,
and $l$ gives the orbital angular momentum in the $Dh^0$ 
system~\cite{dstar}.

The next step is to include the matrix element for the decay 
$D \to \kspipi$.
We follow~\cite{anton} and describe the amplitude for a 
$\bar{D}^0$ decay to this final state as $f(m_+^2,m_-^2)$, 
where $m_+^2$ and $m_-^2$ are the squares of the
two-body invariant masses of the $\ks\pi^+$ and $\ks\pi^-$ combinations.
Assuming no $CP$ violation in the neutral $D$ meson system,
the amplitude for a $D^0$ decay is then given by $f(m_-^2,m_+^2)$.
In the Standard Model, $\left| q/p \right| = 1$ to a good approximation,
and, in the usual phase convention, 
${\rm arg}\left(q/p\right) = 2 \phi_1$.
We then obtain,
\begin{eqnarray}
  \label{eq:m_b0bar}
  M_{\bar{B}^0}(\dt) & = &
  f(m_-^2,m_+^2)  \cos(\dm\dt/2) - 
  i e^{-i2\phi_1} \xi_{h^0} (-1)^l f(m_+^2,m_-^2) \sin(\dm\dt/2), \\
  \label{eq:m_b0}
  M_{B^0}(\dt) & = &
  f(m_+^2,m_-^2) \cos(\dm\dt/2) - 
  i e^{+i2\phi_1} \xi_{h^0} (-1)^l f(m_-^2,m_+^2) \sin(\dm\dt/2).
\end{eqnarray}
The time-dependent Dalitz plot density, $p$, is defined by
\begin{eqnarray}
  \label{eq:pdf}\nonumber
p(m_+^2, m_-^2, \dt) &=& \frac{e^{-|\dt|/\tau_{B^0}}}{4\tau_{B^0}}
\{ 1 + q[ A(m_-^2, m_+^2)\cos(\dm\dt) + 
S(m_-^2, m_+^2)\sin(\dm\dt)]\},\\\nonumber
A(m_-^2, m_+^2)&=&\frac{|f(m_-^2,m_+^2)|^2 - |f(m_+^2,m_-^2)|^2}
{|f(m_-^2,m_+^2)|^2 + |f(m_+^2,m_-^2)|^2},\\
S(m_-^2, m_+^2)&=&\frac{-2\xi_{h^0}(-1)^l 
Im(f(m_-^2,m_+^2)f^*(m_+^2,m_-^2)
e^{+i2\phi_1})}{|f(m_-^2,m_+^2)|^2 + |f(m_+^2,m_-^2)|^2},
\end{eqnarray}
where the $b$-flavor charge is $q = +1$ ($-1$) when the tagging $B$ meson
is a $B^0$ ($\bar{B}^0$).
Thus the phase $2\phi_1$ can be extracted from 
a time-dependent Dalitz plot fit to $B^0$ and $\bar{B}^0$ data.
Note that this formulation assumes that there is no direct $CP$
violation in the $B$ decay amplitudes.


This analysis is based on $386\times 10^6$ $B\bar{B}$ events collected 
with the
Belle detector at the asymmetric energy $e^+e^-$ collider~\cite{KEKB}.
The Belle detector has been described elsewhere~\cite{belle}.
We reconstruct the decays $\bdnpn$ for $h^0 = \pi^0, \eta$ and 
$\omega$ and $\bdspn$ for $h^0 = \pi^0$ and $\eta$.
We use the subdecays $D^{*0}\to D^0\pi^0$, $D^0\to\kspipi$,
$\ks \to \pi^+\pi^-$, 
$\pi^0 \to \gamma\gamma$, $\eta \to \gamma\gamma, \pi^+\pi^-\pi^0$ 
and $\omega \to \pi^+\pi^-\pi^0$.

Charged tracks are selected with a set of requirements based on the
average hit residual and impact parameter relative to the
interaction point (IP).
A transverse momentum of at least 0.1~GeV$/c$ is required for each 
track in order to reduce the combinatorial background. 
All charged tracks that are not positively identified as electrons 
are treated as pion candidates.

Neutral kaons are reconstructed via the decay $\ks\to\pi^+\pi^-$
with no PID requirements for the daughter pions.
The two-pion invariant mass is required to be within 9~MeV$/c^2$
($\sim 3\sigma$) of the $K^0$ mass and the displacement of the 
$\pi^+\pi^-$ vertex from the IP in the transverse ($r$-$\varphi$) 
plane is required to be between 0.2~cm and 20~cm. 
The direction of a vector in the $r-\varphi$ plane from the IP to the 
$\pi^+\pi^-$ vertex is required to agree within 0.2 radians with the 
combined momentum of the two pions.

Photon candidates are selected from calorimeter showers not associated
with charged tracks.
An energy deposition of at least 50~MeV and a photon-like shape are 
required for each candidate.
A pair of photons with an invariant mass 
within 12~MeV$/c^2$ ($2.5\sigma$) of the $\pi^0$ mass is 
considered as a $\pi^0$ candidate.

We reconstruct neutral $D$ mesons in the $\kspipi$ decay channel
and require the invariant mass to be within 15~MeV$/c^2$ ($2.5\sigma$)
of the nominal $\db$ mass.
$\ds$ candidates are reconstructed in the $\db\pi^0$ decay channel.
The mass difference between $D^{*0}$ and $D^0$ candidates is required 
to be within 3~MeV$/c^2$ of the expected value ($\sim 3\sigma$).
$\omega$ candidates are reconstructed in the $\pi^+\pi^-\pi^0$ decay 
channel. Their invariant mass is required to be within 20~MeV$/c^2$ 
($2.5~\Gamma$) of the $\omega$ mass. 
We define the angle $\theta_\omega$ between the normal to the 
$\omega$ decay plane and opposite of $B$ direction in the rest
frame of $\omega$
and require $|\cos \theta_\omega|>0.3$.
We reconstruct  $\eta$ candidates in $\gamma\gamma$ and 
$\pi^+\pi^-\pi^0$ final states and require the invariant mass to be 
within 10 and 30~MeV$/c^2$ ($2.5\sigma$) of the $\eta$ mass, 
respectively. The photon energy threshold for the prompt $\pi^0$ 
and $\eta$ candidates is
increased to 200~MeV in order to reduce combinatorial background.
We remove $\eta$ candidates if either of the daughter photons can be 
combined with any other photon with $E_\gamma>100$~MeV to form a 
$\pi^0$ candidate.

We combine $\db$ and $h^0=\{\pi^0,\omega,\eta\}$ candidates to 
form $B$ mesons. $B$ candidates are also reconstructed from
combinations of $\ds$ and $\pi^0$ or $\eta$ candidates.
Signal candidates are identified by their 
energy difference in the center of mass (CM) system, 
\mbox{$\de=(\sum_iE_i)-E_{\rm beam}$}, and the beam constrained mass, 
$\mbc=\sqrt{E_{\rm beam}^2-(\sum_i\vec{p}_i)^2}$, where $E_{\rm beam}$ 
is the beam energy and $\vec{p}_i$ and $E_i$ are the momenta and 
energies of the decay products of the $B$ meson in the CM frame. 
We select events with $\mbc>5.2$~GeV$/c^2$ and $|\de|<0.3$~GeV,
and define the signal region to be
$5.272$~GeV$/c^2<\mbc<5.287$~GeV$/c^2$, 
$-0.1~{\rm GeV}<\de<0.06$~GeV ($\pi^0$, $\eta\to\gamma\gamma$) 
$|\de|<0.03$~GeV ($\omega$, $\eta\to\pi^+\pi^-\pi^0$).
In cases with more than one candidate in an event, the one with
$\db$ and $h^0$ masses closest to the nominal values
is chosen.

To suppress the large combinatorial background dominated by 
the two-jet-like $e^+e^-\to\qq$ continuum 
process, variables that characterize the event topology are used. 
We require $|\cos\theta_{\rm thr}|<0.80$, where $\theta_{\rm thr}$ is 
the angle between the thrust axis of the $B$ candidate and that of the 
rest of the event. This requirement eliminates 77\% of the continuum 
background and retains 78\% of the signal. We also construct a 
Fisher discriminant, ${\cal F}$, which is based on the production
angle of the $B$ candidate, the angle of the $B$ candidate thrust axis 
with respect to the beam axis, and nine parameters that characterize 
the momentum flow in the event relative to the $B$ candidate thrust 
axis in the CM frame. We impose a requirement on 
${\cal{F}}$ that rejects 67\% of the remaining continuum background 
and retains 83\% of the signal. 

\begin{figure}
\includegraphics[width=0.24\textwidth] {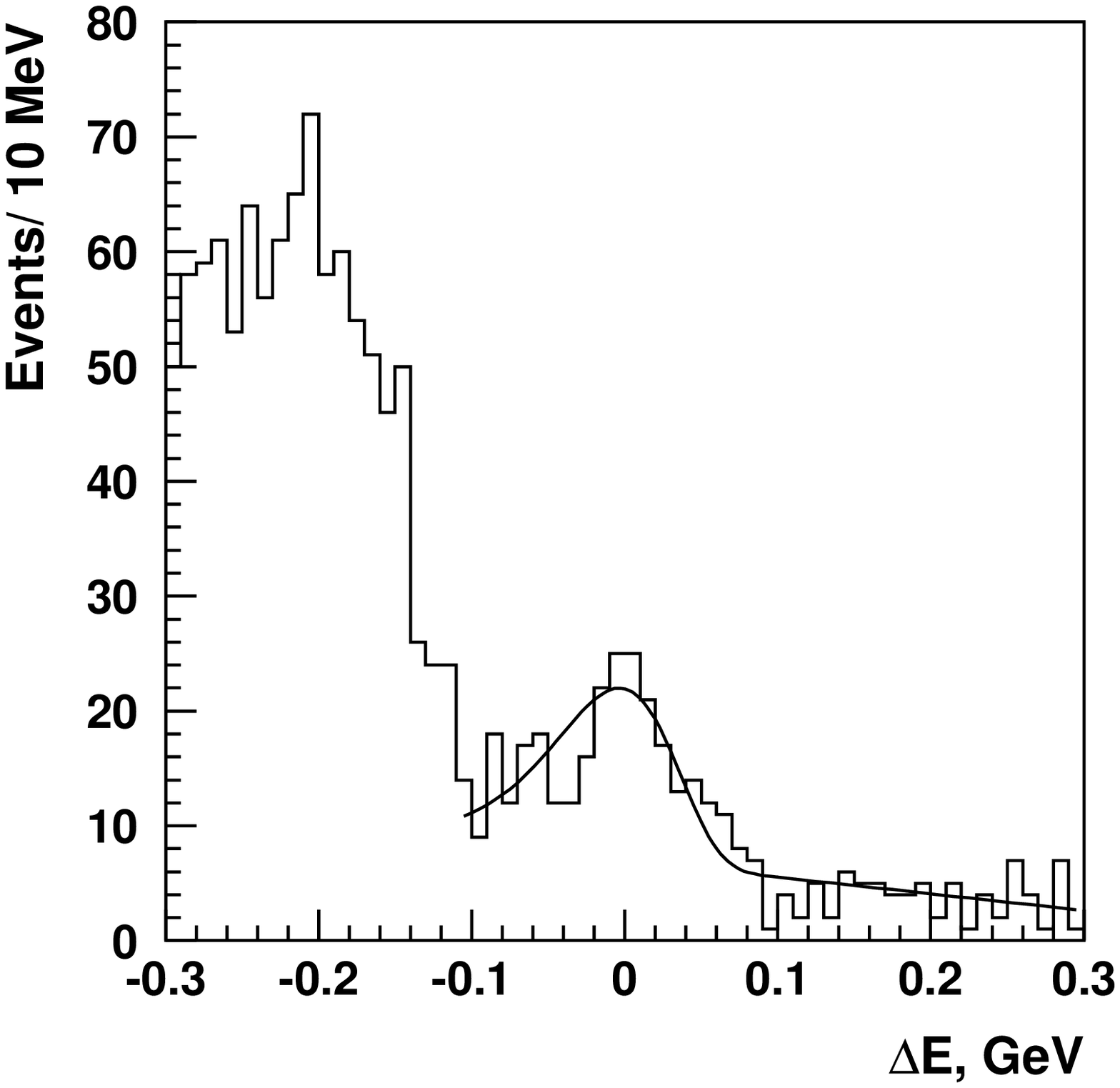}
\includegraphics[width=0.24\textwidth] {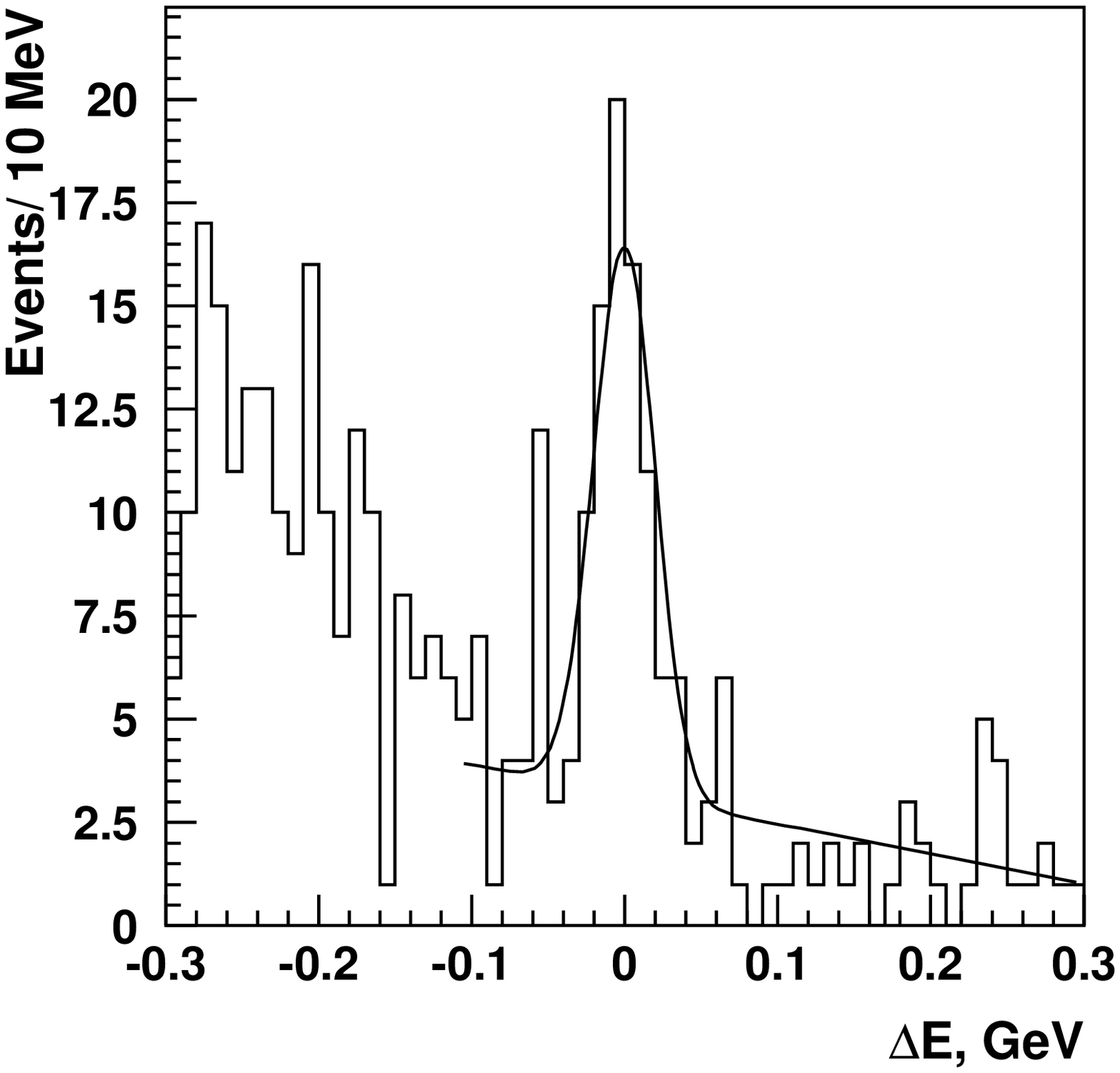}
\includegraphics[width=0.24\textwidth] {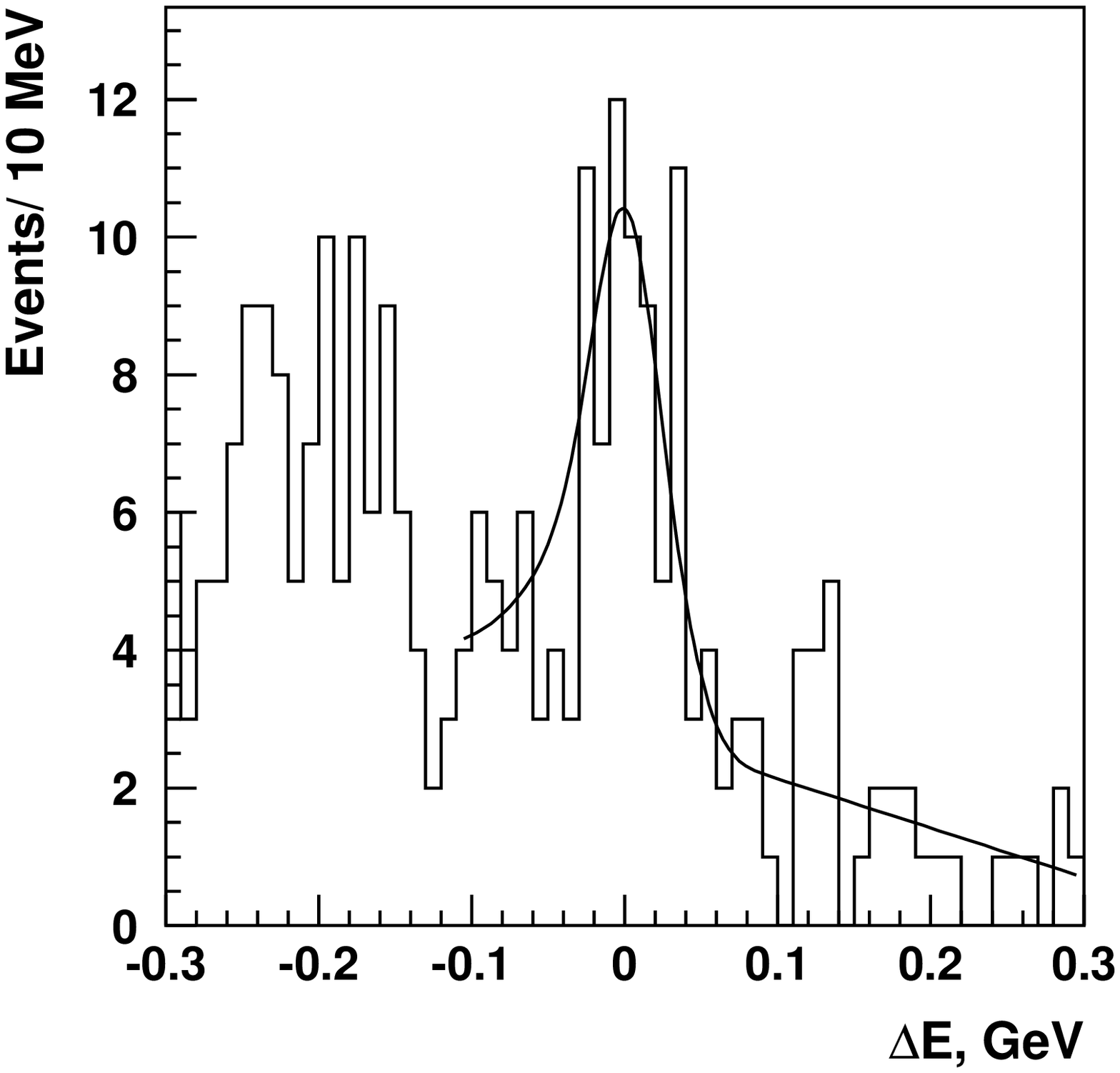}
\includegraphics[width=0.24\textwidth] {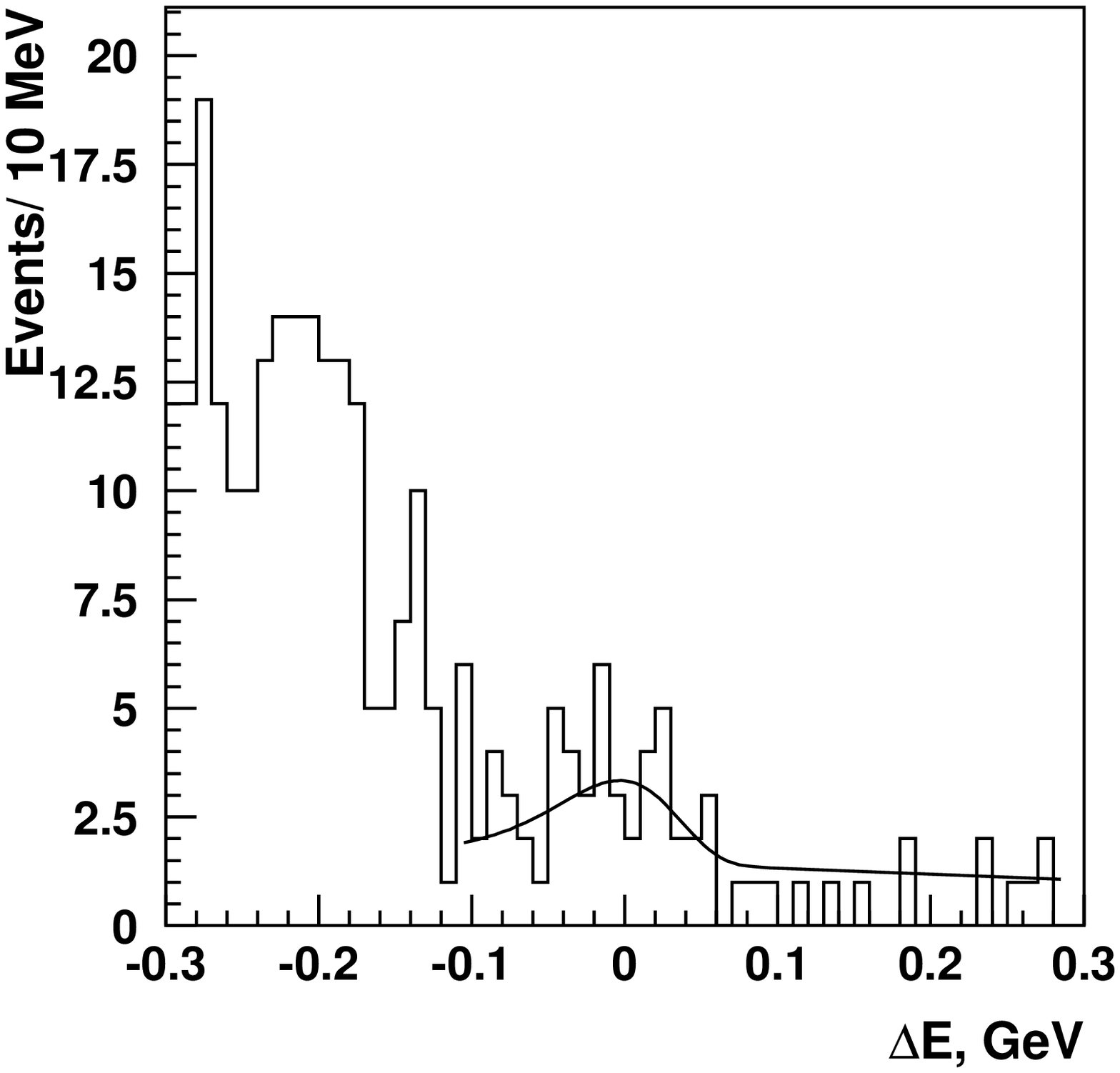}
\includegraphics[width=0.24\textwidth] {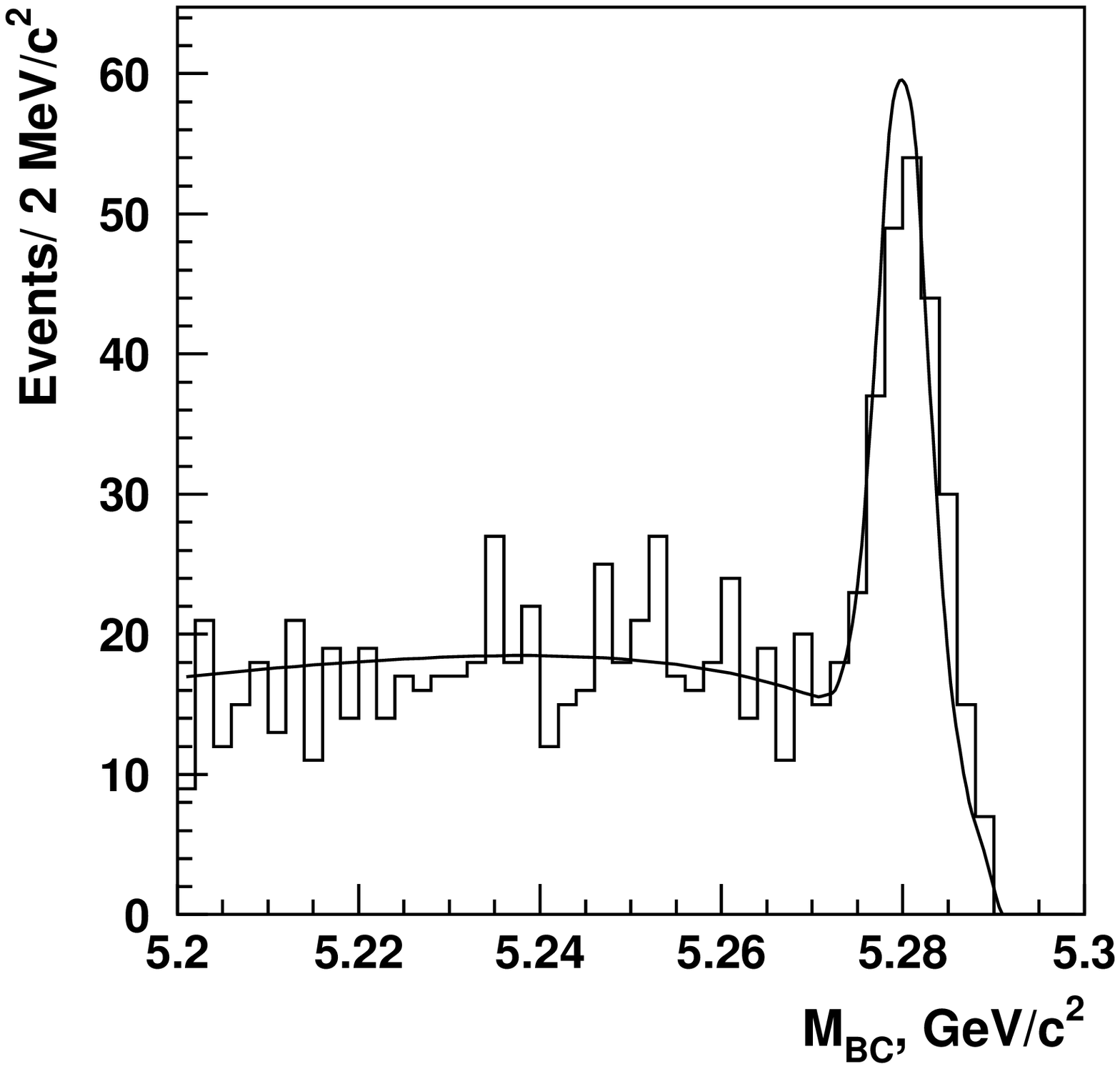}
\includegraphics[width=0.24\textwidth] {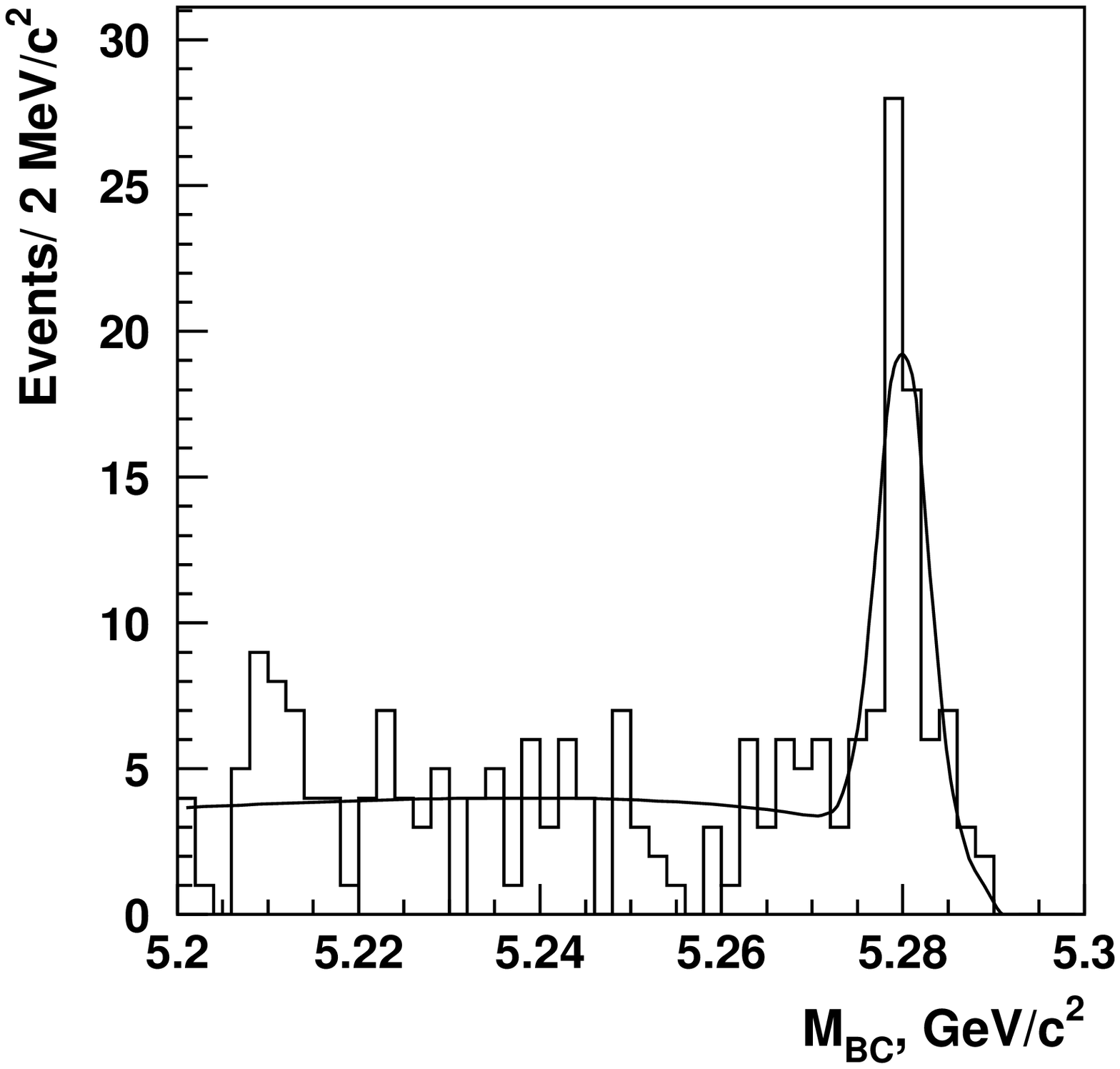}
\includegraphics[width=0.24\textwidth] {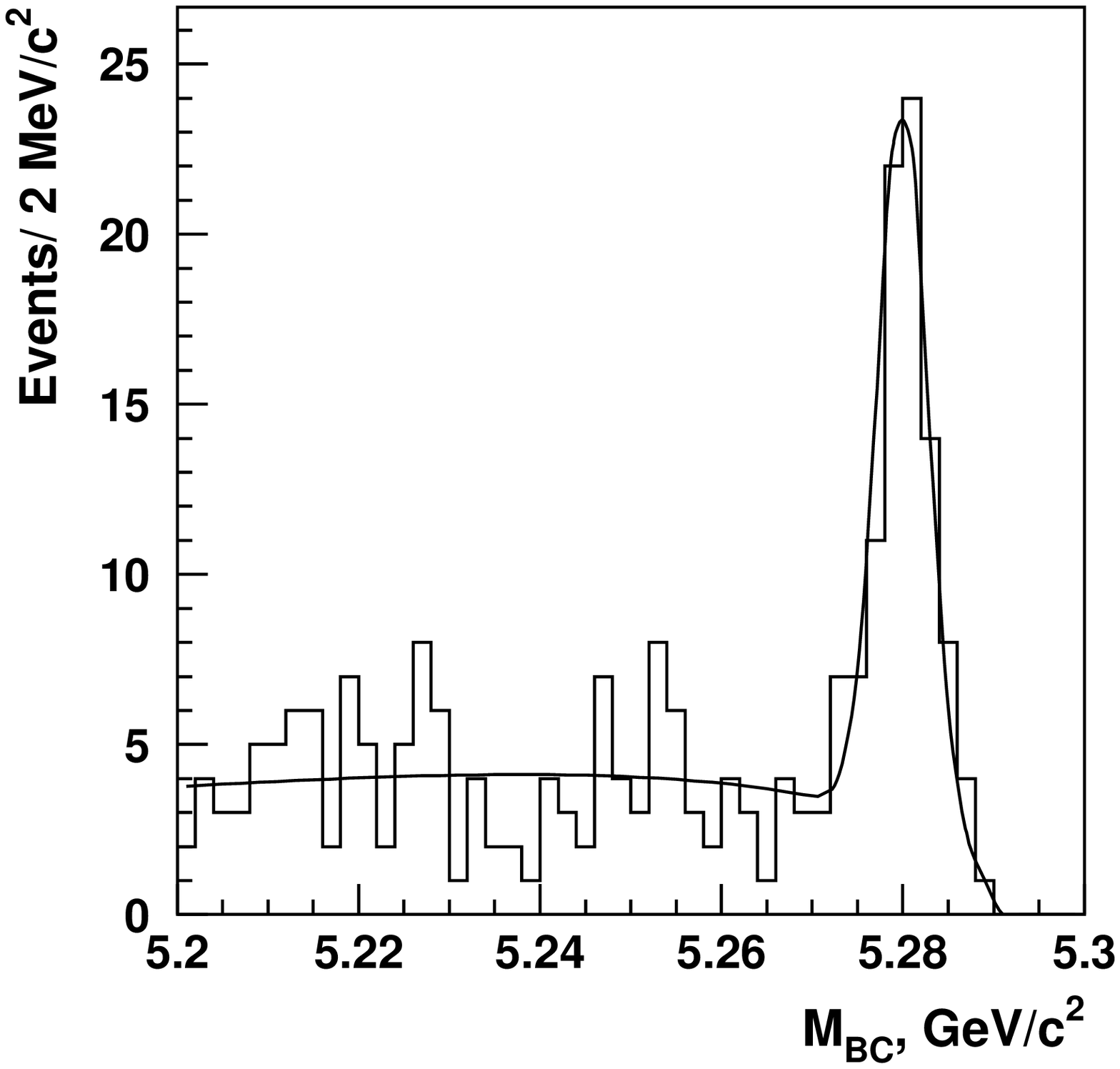}
\includegraphics[width=0.24\textwidth] {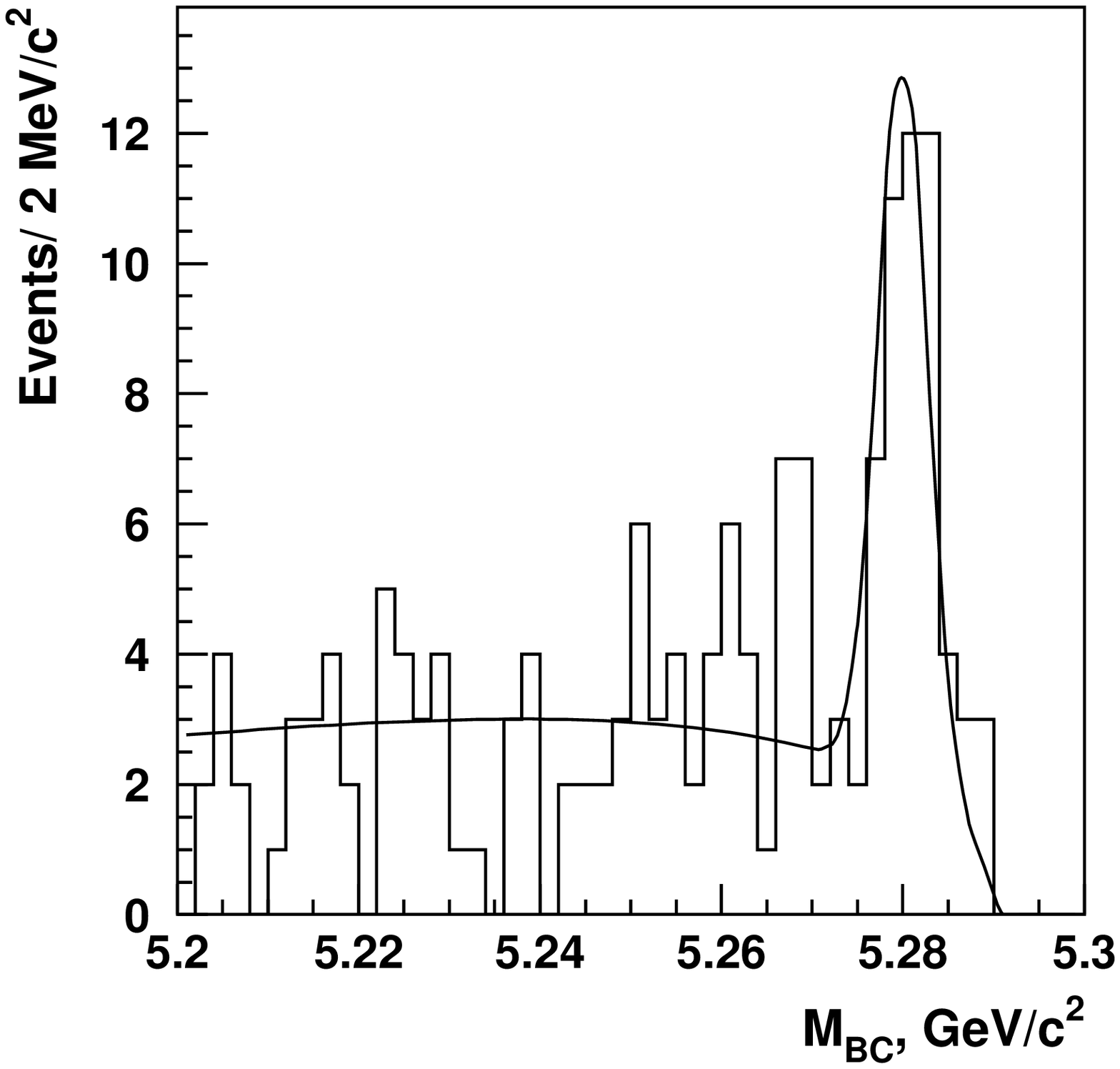}
\caption{From left to right: $\de$ (top) and $\mbc$ (bottom) 
distributions for the $\bar{B}^0$ decays to $D\pi^0$, $D\omega$,
$D\eta$ and $D^*\pi^0$, $D^*\eta$. 
Histograms
represent the data 
and curves shows the results of the fit.}
\label{de_exp}
\end{figure}

Figure~\ref{de_exp} shows the $\de$ and 
$\mbc$ distributions for the events in the signal region.
For each mode, the $\de$ distribution is fitted with a 
Gaussian for signal and a linear function for background. The Gaussian 
mean value and width are fixed to the values from MC simulation of
the signal events. The region $\de<-0.1$~GeV is excluded from the fit 
to avoid contributions from other $B$ decays, such as 
$B\to D h^0 (\pi)$ where $(\pi)$ denotes a possible additional pion.
For the $\mbc$ distribution fit we use the sum of a signal Gaussian
and an empirical background function with a kinematic 
threshold, with a shape fixed from the analysis of the 
off-resonance data.
The signal yield is obtained from the fit to the $\de$ distribution.
The results of these fits to data are summarized in
Table~\ref{tabeff}. 
\begin{table}
  \caption{
    \label{tabeff}
    Number of events in the signal region ($N_{\rm tot}$),
    detection efficiency, number of signal events from the
    $\de$ fit ($N_{\rm sig}$)
    and signal purity for the $B\to D^{(*)} h^0$ final states.
  }
  \vspace{0.5\baselineskip}
  \begin{tabular}
    {|l|c|@{\hspace{3mm}}c@{\hspace{3mm}}|@{\hspace{3mm}}c@{\hspace{3mm}}|c|}
    \hline
    Process & $N_{\rm tot}$& Efficiency (\%) & $N_{\rm sig}$  & Purity \\ 
    \hline
    $D\pi^0$ & 265& 8.7 & $157\pm 24$ & 59\% \\
    $D\omega$& 78 & 4.1 &  $67\pm 10$ & 86\% \\
    $D\eta$  & 97 & 3.9 &  $58\pm 13$ & 60\% \\
$D^*\pi^0$,$D^*\eta$ & 52 &  &  $27\pm 11$ & 52\% \\
    \hline
    Sum      & 492 &     & $309\pm 31$ & 63\% \\
    \hline
  \end{tabular}
\end{table}

The signal $B$ meson decay vertex is reconstructed using the $D$
trajectory and an IP constraint. 
The tagging $B$ vertex position is obtained with the IP constraint and 
with well reconstructed tracks that are not assigned to the signal $B$
candidate.
The algorithm is described in detail elsewhere~\cite{resol}.
The time difference between signal and tagging $B$ candidates are
calculated using $\Delta t=\Delta z/\gamma\beta c$ and 
$\Delta z = z_{CP}-z_{\rm tag}$.


The proper-time interval resolution function $R_\mathrm{sig}(\dt)$
is formed by convolving four components:
the detector resolutions for $z_{CP}$ and $z_\mathrm{tag}$,
the shift in the $z_\mathrm{tag}$ vertex position
due to secondary tracks originating from charmed particle decays,
and the kinematic approximation that the $B$ mesons are
at rest in the CM frame~\cite{resol}.
A small component of broad outliers in the $\Delta z$ distribution,
caused by mis-reconstruction, is represented by a Gaussian function.

Charged leptons, pions, kaons, and $\Lambda$ baryons
that are not associated with a reconstructed $\bdbkspipi$ decay
are used to identify the $b$-flavor of the accompanying $B$ meson.
The tagging algorithm is described in detail 
elsewhere~\cite{flavor}.
We use two parameters, $\fq$ and $r$, to represent the tagging 
information.
The first, $q$, has the discrete value $+1$~($-1$)
when the tag-side $B$ meson is more likely to be a $B^0$~($\bar{B}^0$).
The parameter $r$ corresponds to an event-by-event 
flavor-tagging dilution that ranges
from $r=0$ for no flavor discrimination
to $r=1$ for an unambiguous flavor assignment.

We perform an unbinned time-dependent Dalitz plot fit.
The negative logarithm of the unbinned likelihood function is minimized:
\begin{equation}
  -2 \log L = 
  -2 
  \left[ 
    \sum \limits^n_{i=1} \log p(m^2_{+i}, m^2_{-i}, \dt_i) - 
    \log \int p(m^2_+, m^2_-, \dt) dm^2_+ dm^2_- d\dt 
  \right], 
\end{equation}
where $n$ is the number of events, $m^2_{+i}$, $m^2_{-i}$ and $\dt_i$ 
are the measured invariant masses of the $D$ daughters,
and the time difference between signal and tagging $B$ meson decays,
respectively. 
The function $p(m^2_+, m^2_-, \dt)$ is the time-dependent Dalitz plot 
density, which is calculated according to Eq.~(\ref{eq:pdf})
and incorporates reconstruction efficiency,
flavor-tagging efficiency, wrong tagging probability, background
and $\dt$ resolution.

\begin{figure}
\includegraphics[width=0.45\textwidth] {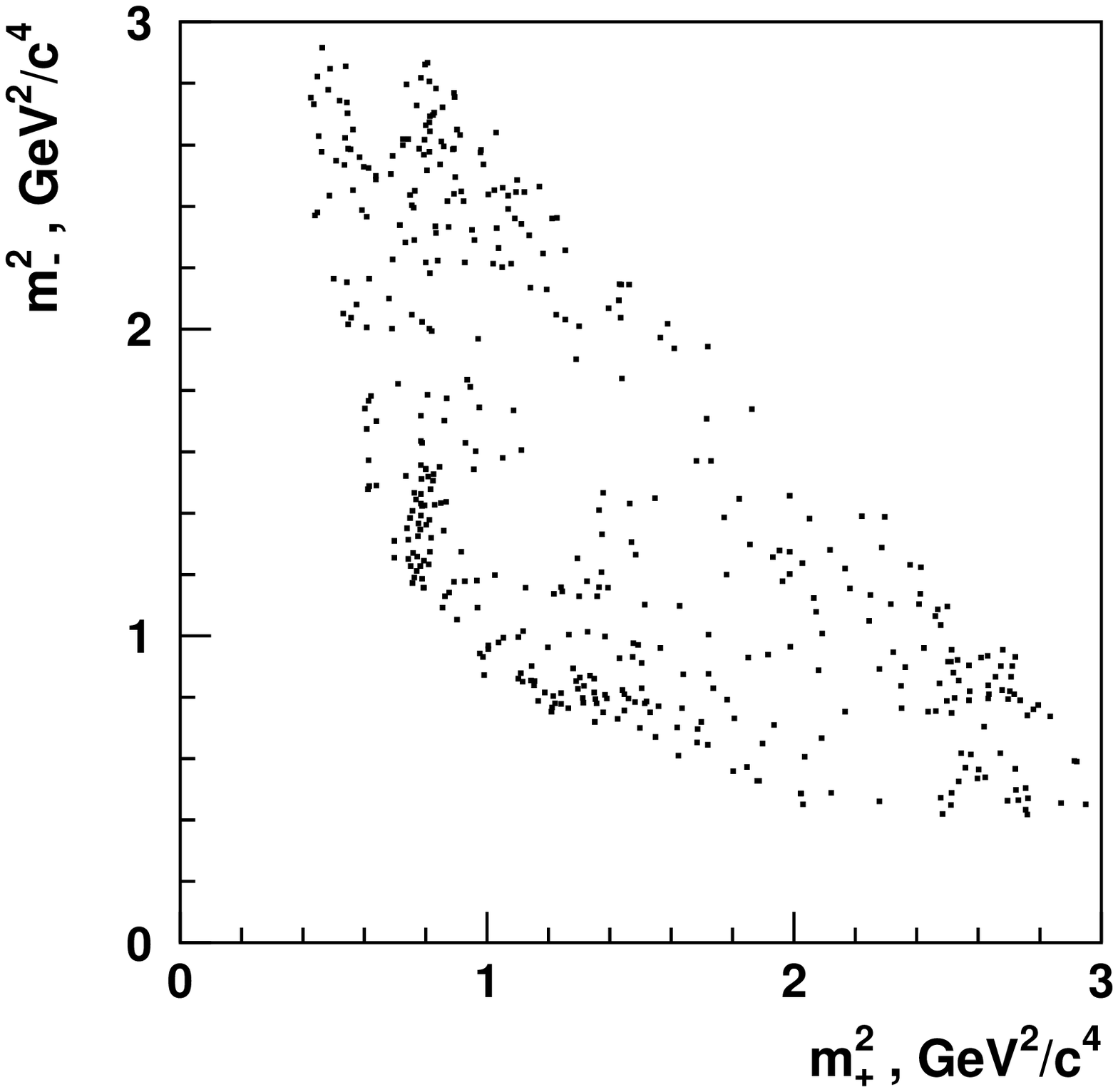}\hfill
\includegraphics[width=0.45\textwidth] {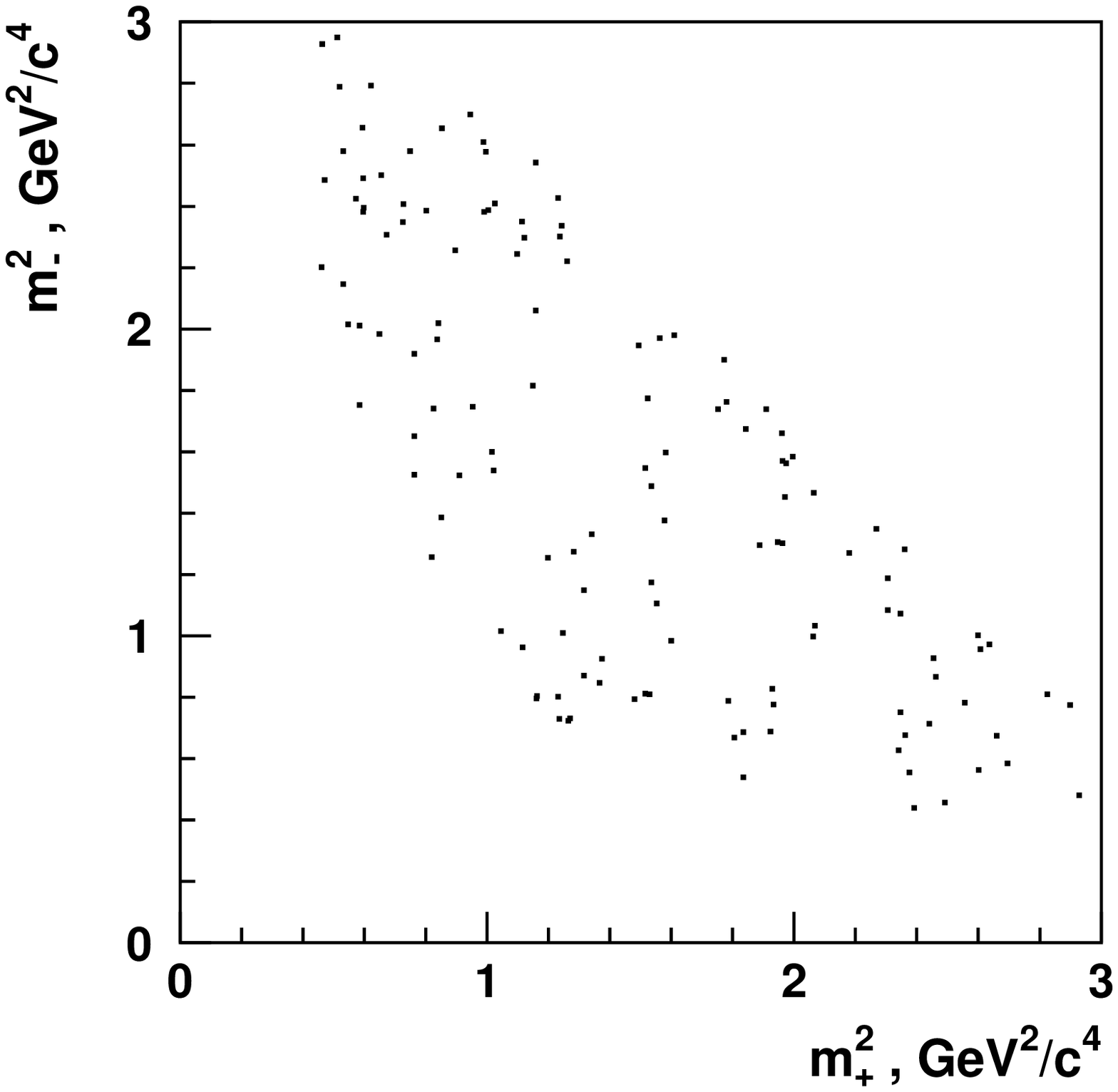}
\caption{Dalitz plot distribution for the $D h^0$ candidates from
  $B$ signal box (left) and $\mbc$ sideband.
$M^2_\pm$ denote the square of $\ks\pi^\pm$ invariant mass.}
\label{dalitz}
\end{figure}

\begin{figure}
\includegraphics[width=0.45\textwidth] {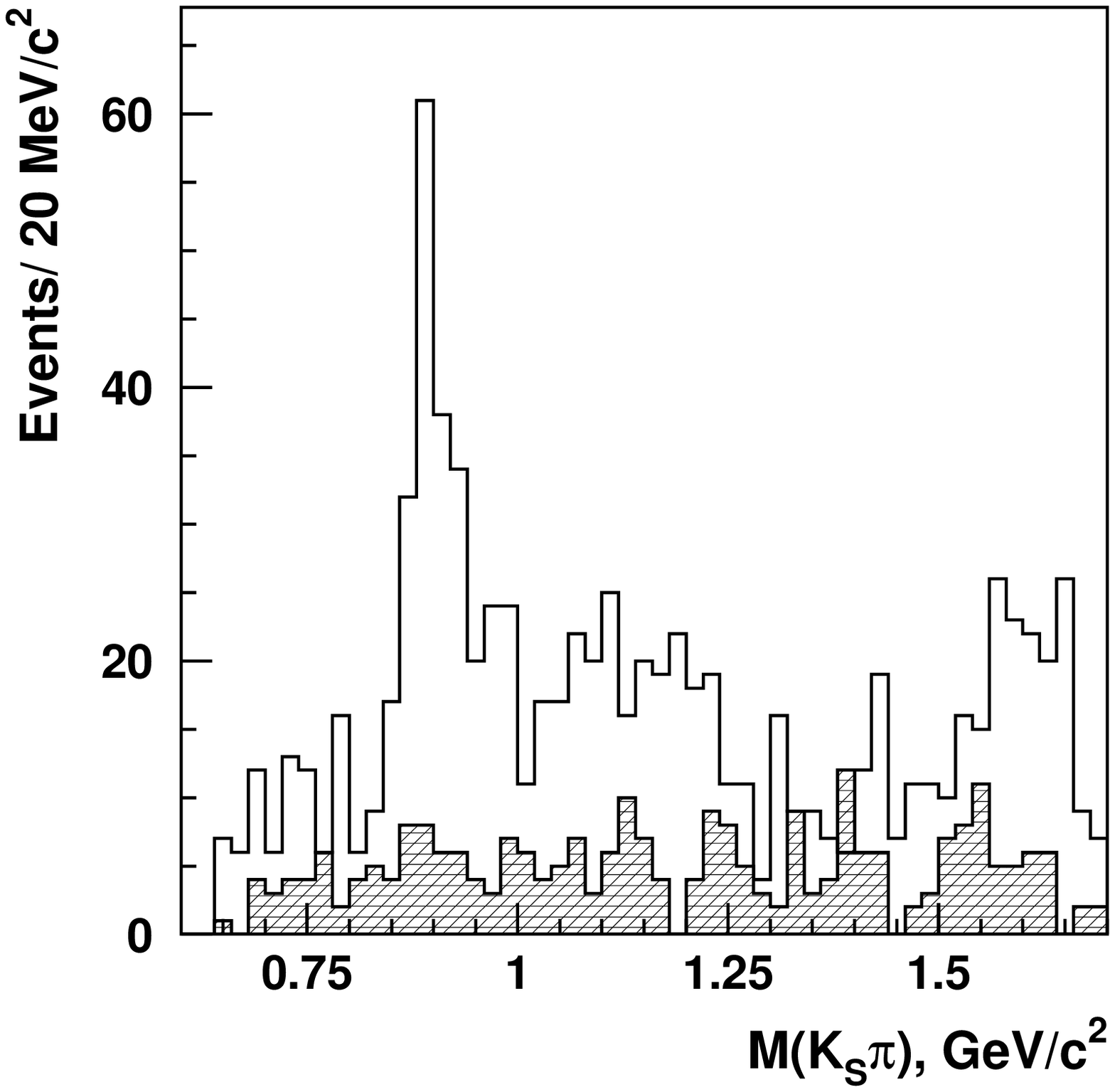}\hfill
\includegraphics[width=0.45\textwidth] {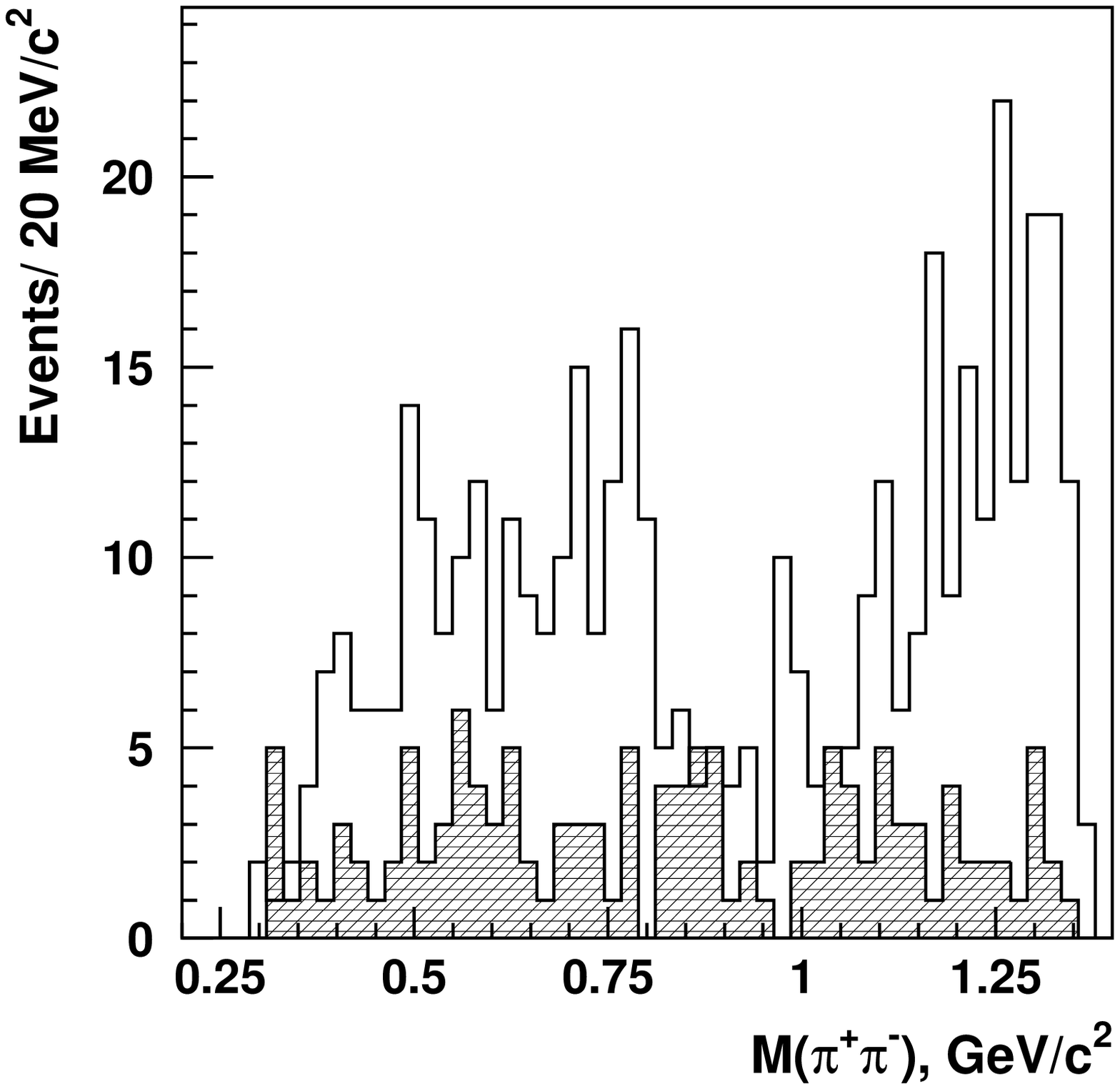}
\caption{$\ks\pi^\pm$ (left) and $\pi^+\pi^-$ (right) invariant mass 
distributions for the $D h^0$ candidates. Open histograms correspond
to the $B$ signal region, hatched ones --- to the $\mbc$ sideband.}
\label{minv}
\end{figure}

We describe the background by the sum of four components:
a) $B$ decays containing real $D$ mesons, 
b) $B$ decays with combinatorial $D$ mesons,
c) $\qq$ events containing real $D$ mesons, 
d) $\qq$ events with combinatorial $D$ mesons. 
The Dalitz plot is described by the function $f(m_+^2,m_-^2)$ for 
a) and c) and by the sum of phase space and $K^*(892)$ contributions
for b) and d). 
The PDF for b) and d) is obtained from an analysis of the events in the 
$\mbc$-$\de$ sidebands.
The $\dt$ distribution for the $B$ decay backgrounds are described by
an exponential convolved with the detector resolution. For the $\qq$
background, a triple Gaussian form is used, which is obtained from 
events with $|\cos\theta_{\rm thrust}|>0.8$.
We use the experimental data and generic MC to fix the fractions of
background components.
Figure~\ref{dalitz} shows the Dalitz plot distributions for the 
signal and background events, integrated over the entire $\dt$ range 
and $B^0$ and $\bar{B}^0$ combined.
Projections of the signal candidate distribution on the 
$M(\ks\pi^\pm)$ (two entries per event)
and $M(\pi^+\pi^-)$ axes are shown in Fig.~\ref{minv}.

The procedure for the $\dt$ measurement is tested by extracting 
$\tau_{B^+}$ using $B^+\to \bar{D}^0[\kspipi]\pi^+$ decay. 
We obtain $\tau_{B^+}=1.678\pm 0.043$~ps (statistical error only),
consistent with the PDG~\cite{PDG} value $1.638\pm 0.011$~ps. 

\begin{figure}
  \includegraphics[width=0.41\textwidth]{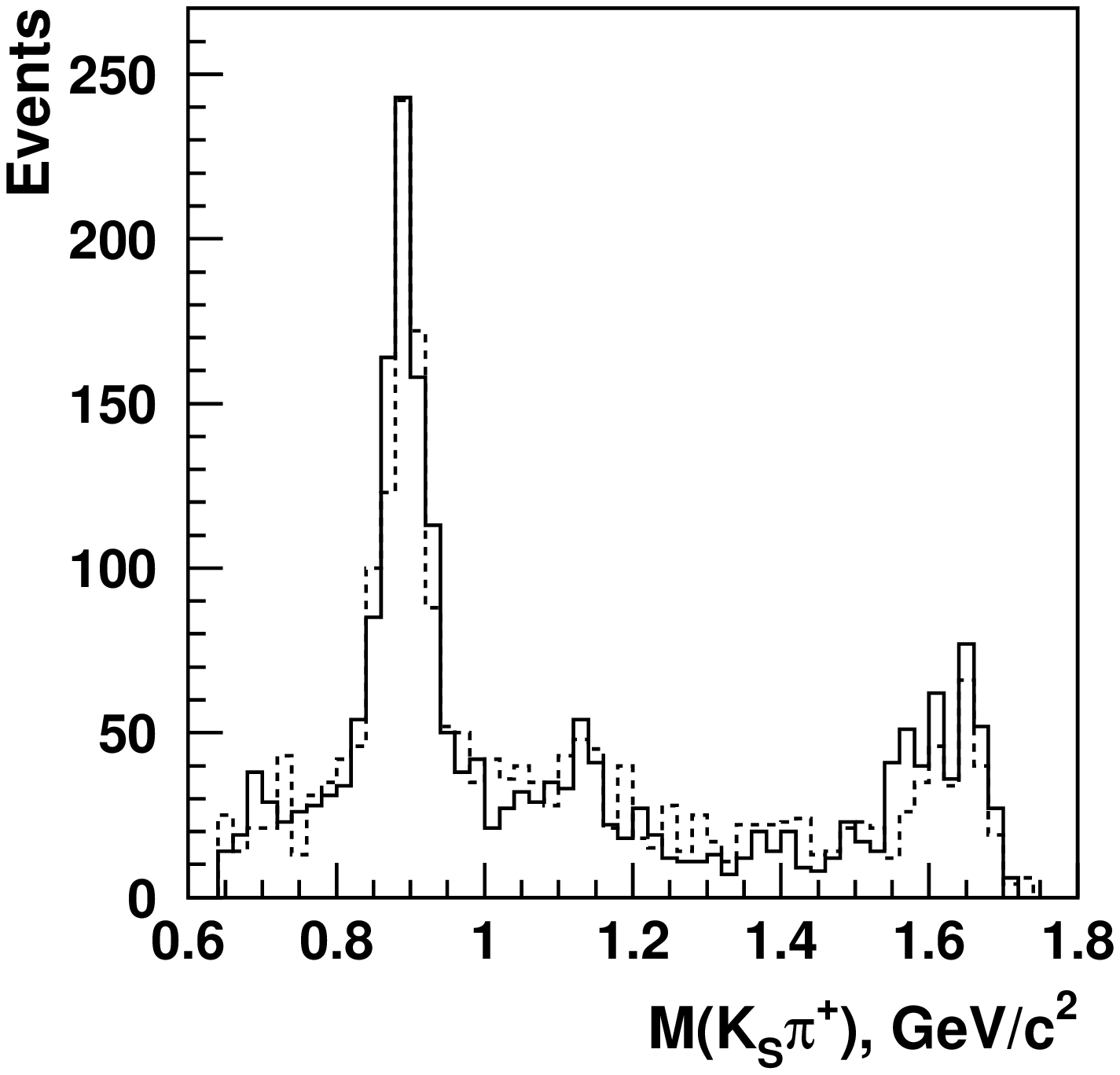}
  \includegraphics[width=0.41\textwidth]{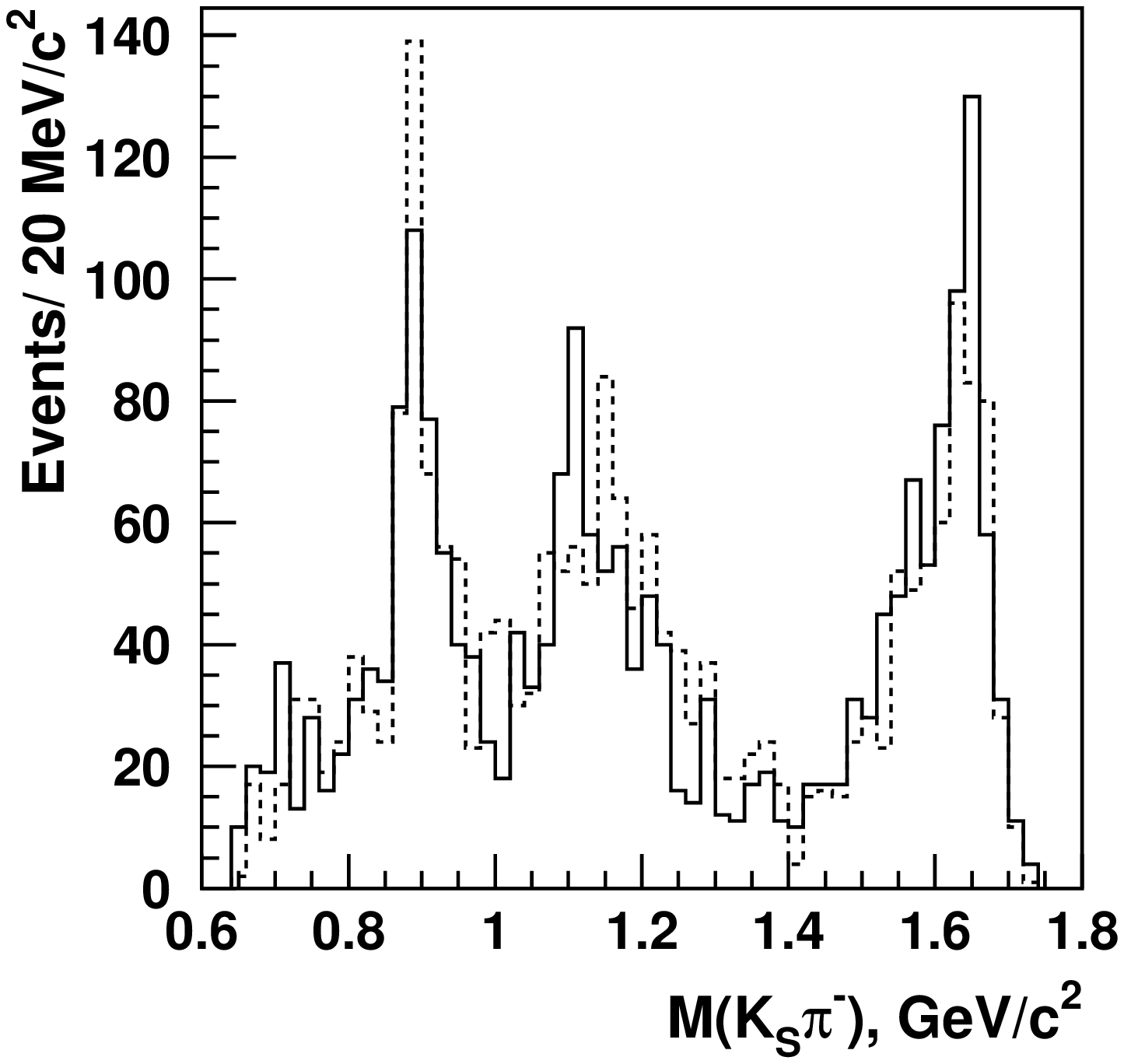}
  \includegraphics[width=0.41\textwidth]{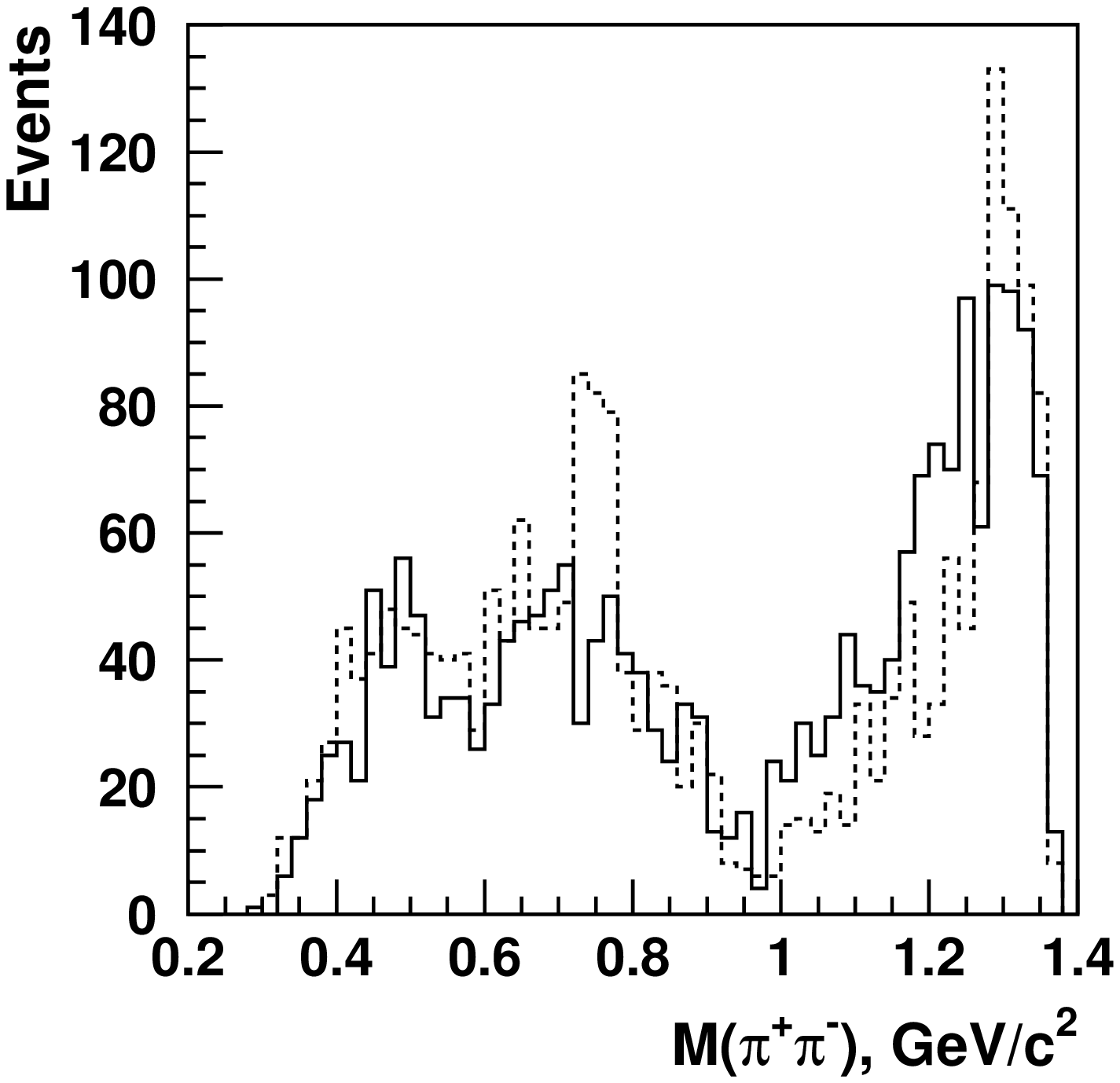}
  \caption{
    \label{mc_dalitz}
MC invariant mass distributions of $D$ decay daughters
produced in the $\bdbkspipi$ decay chain. Events are generated with 
$2\phi_1=47^\circ$.
Only events with good tagging quality, $r>0.5$, are shown.
The dashed histograms show distributions for 
events with $q\cdot\dt<-\tau_{B^0}/2$, 
the solid histograms show those for 
events with $q\cdot\dt>\tau_{B^0}/2$.
This range is chosen to enhance the visible asymmetry.}
\end{figure}

\begin{figure}
  \includegraphics[width=0.41\textwidth]{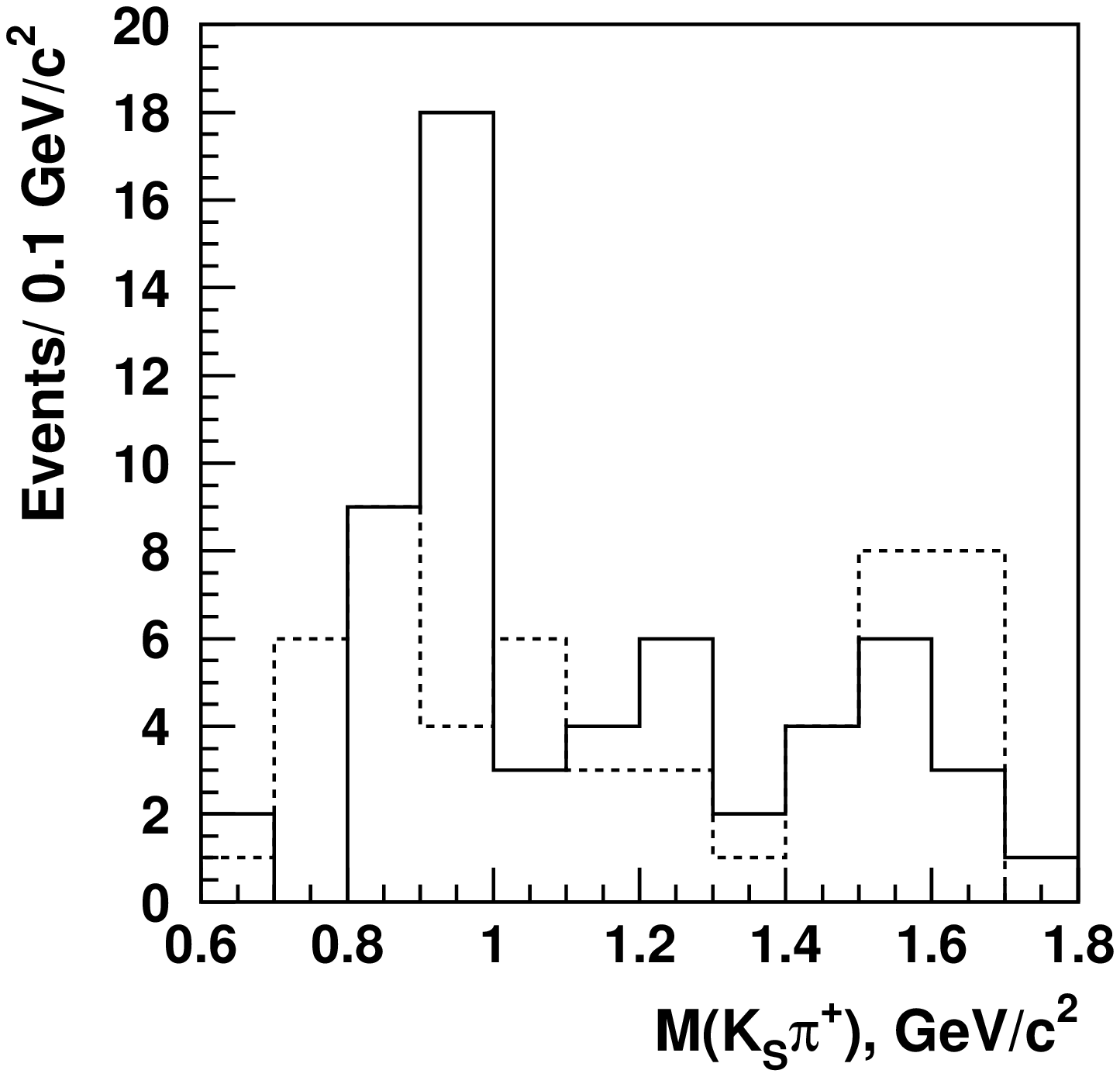}
  \includegraphics[width=0.41\textwidth]{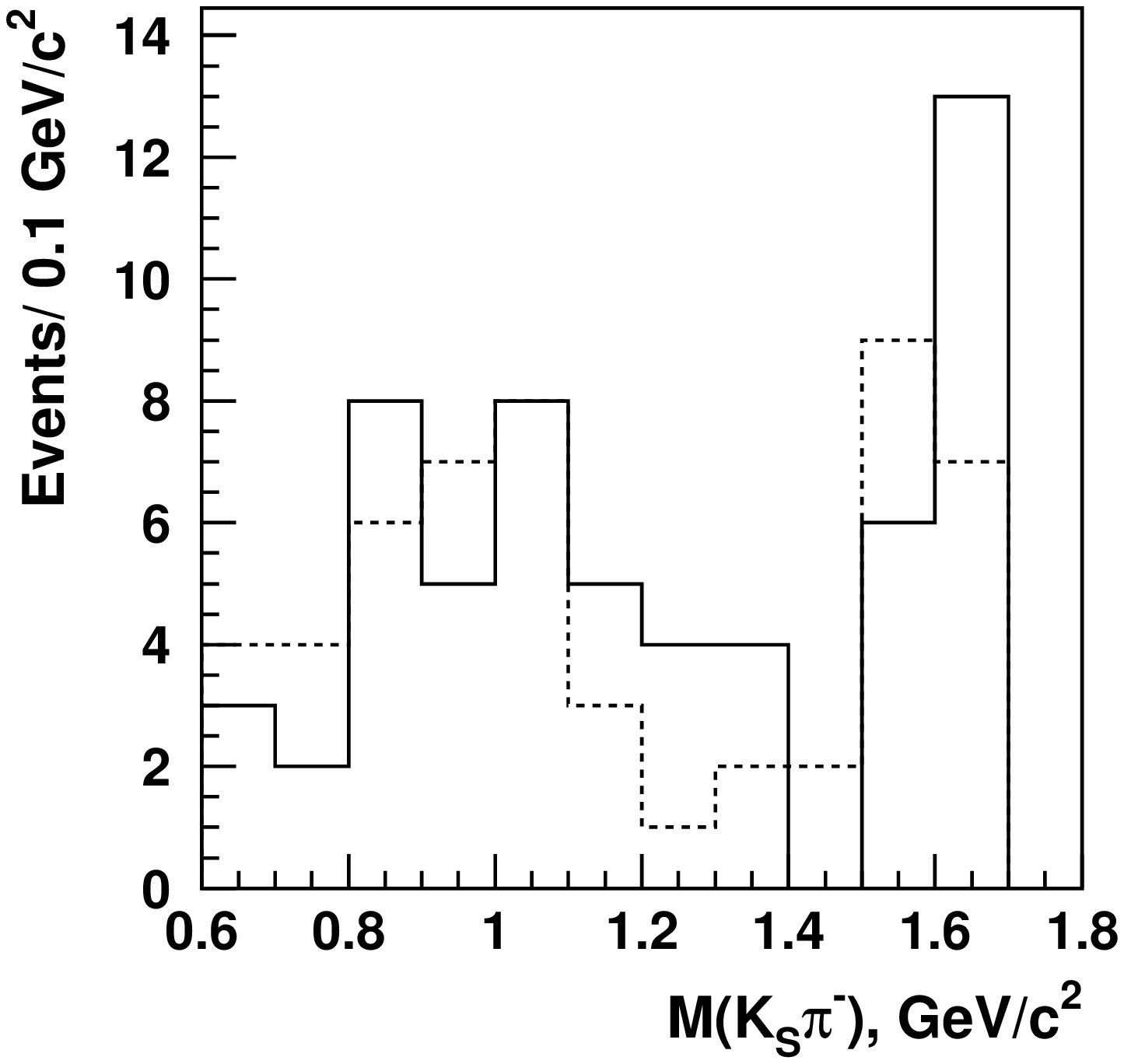}
  \includegraphics[width=0.41\textwidth]{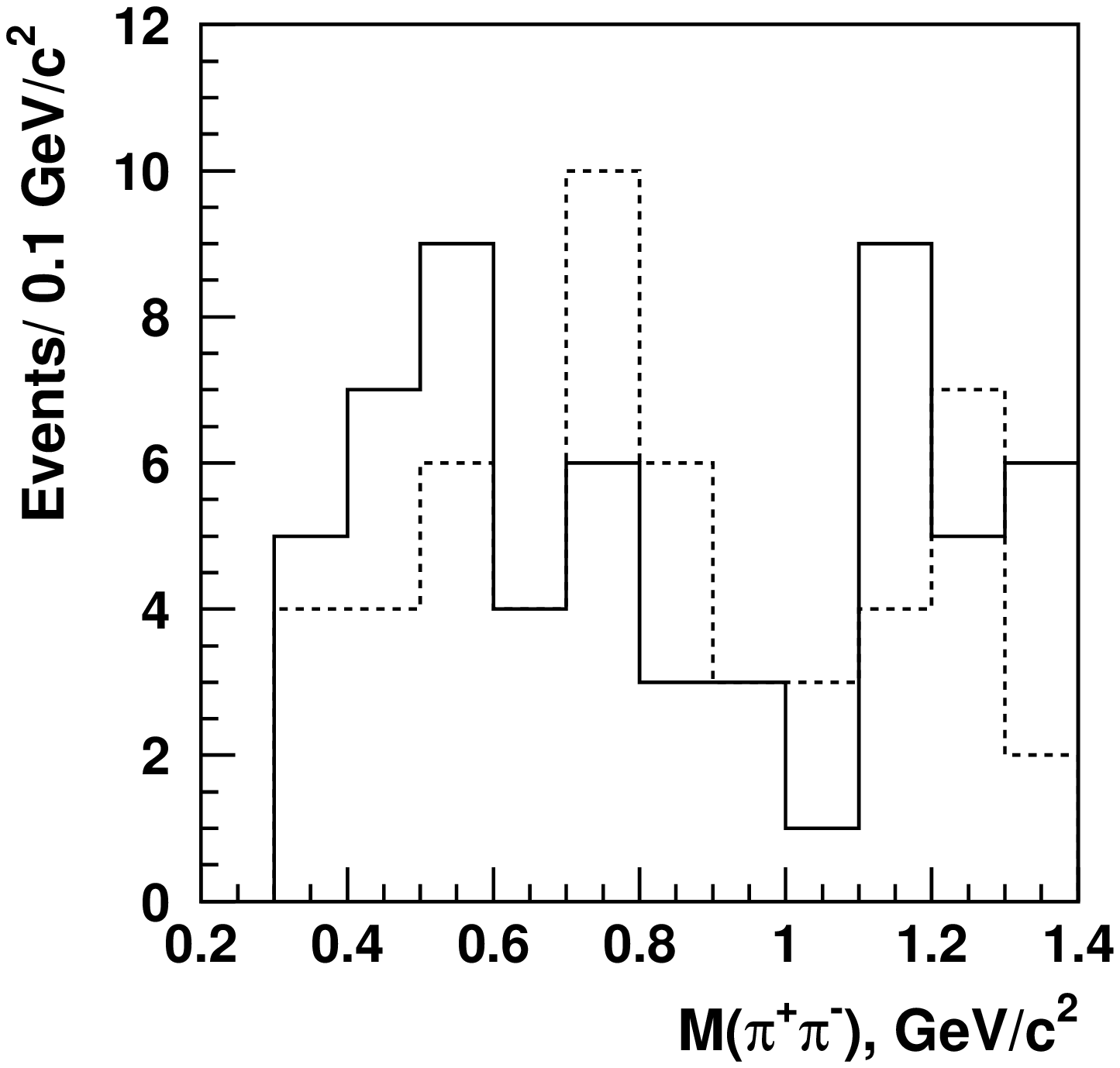}
  \caption{
    \label{exp_dalitz}
Invariant mass distributions of $D$ decay daughters for the 
$B\to D[\kspipi] h^0$ candidates. 
Only events with good tagging quality, $r>0.5$, are shown.
The dashed histograms show distributions for 
events with $q\cdot\dt<-\tau_{B^0}/2$, 
the solid histograms show those for 
events with $q\cdot\dt>\tau_{B^0}/2$.}
\end{figure}

The potential accuracy of the $\phi_1$ determination is 
estimated using a Monte Carlo study.
We generate $\bdbkspipi$ decays using $2\phi_1 = 47^\circ$
and process the events
with detector simulation, reconstruction, flavor tagging and the $CP$ fit.
In Fig.~\ref{mc_dalitz} we show the invariant mass distributions
of the $D$ decay daughters 
for MC events with $q\cdot\dt$ greater than $\tau_{B^0}/2$ and
for events with $q\cdot\dt$ less than $-\tau_{B^0}/2$. 
This range is chosen to enhance the visible asymmetry.
Events with $|\dt| < \tau_{B^0}/2$ 
are not shown.
The MC statistics correspond to about 30 times the size of the data.
We see clear differences in the two invariant mass distributions due
to $CP$ violation.
Figure~\ref{exp_dalitz} shows the corresponding distributions for the 
data.

We divide the $\phi_1=[0^\circ:180^\circ]$ range into 18 points in 
steps of $10^\circ$. 
For each point we perform 30 pseudo-experiments with data samples
consisting of 300 reconstructed $D\pi^0$ events. 
We add 180 background events to each sample,
where the background is modelled by the shape used in the data analysis.
For each pseudo-experiment,
we perform an unbinned time-dependent Dalitz plot fit.

Thus, for each input value of $\phi_1$ 
we obtain fitted results from 30 pseudo-experiments.
From the means and widths of the distributions of these results 
we obtain average $\phi_1$ fit results and 
estimates of their statistical uncertainties.
We find the fit results are in good agreement with the input values,
and the expected uncertainty on $\phi_1$ is approximately $21^\circ$.

We also study larger ensembles of pseudo-experiments 
for two $\phi_1$ input values: 
$23.5^\circ$ and $66.5^\circ$, which correspond to $\sinphi$. 
We verify that the fit value is unbiased.
The MC pseudo experiments show that the nominal errors from the fit are 
underestimated at the current level of statistics. 
We obtain a $\phi_1$ error 
of $21^\circ$ from the RMS width of the distribution of fit results, 
while the errors from the fit have an average value of $15^\circ$. 
With higher statistics, corresponding to data samples of 1 ab$^{-1}$ 
and greater, these two values are consistent.
Therefore we use the uncertainty from the MC pseudo experiment fit 
study, $21^\circ$, as the statistical error for our measurement.
The distribution of MC pseudo experiment fit results is consistent 
with, and
assumed to be, a Gaussian.  This study also shows that, if a central 
value corresponding to one solution of $\phi_1$ is obtained, the
second solution can be ruled out at 95\% confidence level with our 
current statistics.

We have tested for a possible bias in the method due to neglect
of the suppressed amplitudes.
Due to the smallness of the suppressed amplitude compared to the
$B^0$-$\bar{B}^0$ mixing effect, such bias is expected to be small,
and indeed we find it to be less than $1\%$.

\begin{figure}
\includegraphics[width=0.45\textwidth] {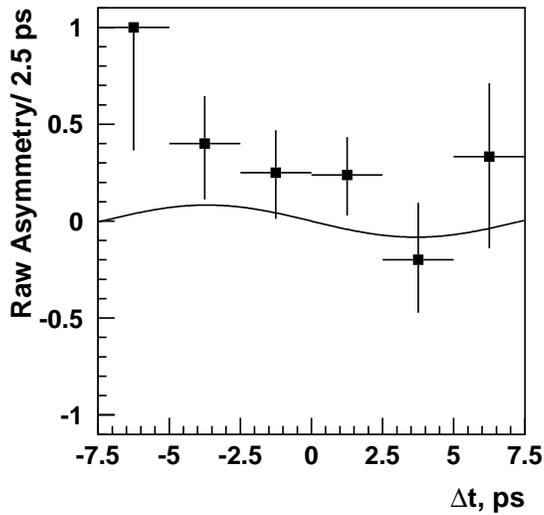}
\caption{Raw asymmetry distribution for the $D[\ks\rho^0]h^0$ 
candidates. The smooth curve is the result of the fit to the full
Dalitz plot.}
\label{asym}
\end{figure}

We perform a fit by fixing $\tau_{B^0}$ and $\dm$ at the PDG 
values, using a fixed background shape as described above, and using
$\phi_1$ and the background fraction as fitting parameters.
The $de$ fit results are used to constrain the background fraction.
The results are given in Table~\ref{fitres} for each of the three final
states separately and for the simultaneous fit over all modes.
Errors are statistical only and  determined from the MC pseudo 
experiments.
\begin{table}
  \caption{
    \label{fitres}
    Fit results for the data. The statistical errors shown here are
  determined from the MC pseudo experiments.
  }
  \vspace{0.5\baselineskip}
  \begin{tabular}
    {|l|c|}
    \hline\hline
Final state & $\phi_1$ fit result, $^\circ$\\ \hline
$D\pi^0$  & $11\pm 26$ \\
$D\omega$, $D\eta$ & $28\pm 32$ \\ 
$D^*\pi^0$,$D^*\eta$ & $25\pm 35$\\\hline
Simultaneous fit & $16\pm 21$\\\hline

    \hline
  \end{tabular}
\end{table}
To illustrate the asymmetry, in Fig.~\ref{asym} we show the raw 
asymmetry distribution for the $Dh^0$ candidates, with an 
additional constraint to select events consistent with $D\to\ks\rho$: 
$|M_{\pi^+\pi^-}-0.77 {\rm GeV/c^2}|<0.15$~ GeV$/c^2$. 
In this case the system
behaves as a $CP$ eigenstate, with an asymmetry proportional to
$-\sin 2\phi_1$.  We do not include $D^*h^0$ candidates in 
Fig.~\ref{asym} since these have the opposite asymmetry.
The curve corresponds to the value $\phi_1=16^\circ$ found in the
simultaneous time-dependent Dalitz plot fit over all data.

The model used for the $\dnkspipi$ decay is one of the main sources of 
systematic error for our analysis. 
A MC simulation is used to estimate the effects of the model 
uncertainties. 
We use three models to estimate this error:
the first is a simple model from the CLEO analysis of $\dnkspipi$ 
decay~\cite{dkpp_cleo}, the other two are from a similar Belle 
analysis~\cite{anton,anton_dkstar}.
Event samples are generated according to each model, then we perform 
a fit with the model assumption from~\cite{anton}. The difference 
between the input $\phi_1$ value and fit result does not exceed 5$^\circ$.
We use different background descriptions to measure the systematic
uncertainty due to the background parameterization.
Several models are used for the Dalitz plot distribution of the 
background: 
only a uniform distribution and a signal $D$ PDF are compared.
For the time dependence, we consider cases with only a $\qq$ component
or only a $\bb$ component.
The difference in $\phi_1$ results for these models does not exceed 
10$^\circ$.
The error due to the vertexing and flavor tag is
similar to those in other $CP$ analyses~\cite{sin2phi1} (about 1$^\circ$). 

We have presented a new method to measure the Unitarity Triangle angle 
$\phi_1$ using a time-dependent amplitude analysis of the 
$D\to\kspipi$ decay produced in the processes 
$\bar{B}^0 \to Dh^0$. 
We find $\phires$. The first error is statistical and determined 
from MC studies. The second is systematic.
The 95\% CL region including systematic uncertainty is
$\phicl$, thus ruling out the second solution from
$\sinphi$ at 97\% CL.

\end{document}